\documentclass{PoS}

\usepackage{amsmath,amssymb}
\usepackage{graphicx,bm}

\allowdisplaybreaks

\title{Dense baryonic matter in the hidden local symmetry approach}

\ShortTitle{Dense baryonic matter in the hidden local symmetry approach}

\author{\speaker{Yong-Liang Ma}\thanks{A footnote may follow.}\\
        Department of Physics,  Nagoya University, Nagoya, 464-8602, Japan\\
        E-mail: \email{ylma@hken.phys.nagoya-u.ac.jp}}

\abstract{
The dense baryonic matter and hadron properties in medium were simulated by using a Skyrme model including the ground state vector mesons introduced from the hidden local symmetry approach up to the next to leading order. We found that both the $\rho$ and $\omega$ mesons affect the baryonic matter and medium modified hadron properties dramatically. The most remarkable observation is that, the pion decay constant and the nucleon mass have the similar density dependence which agrees with the large $N_c$ argument. Explicitly, they drop with increasing density in the Skyrmion phase and stop decreasing at $n_{1/2}^{}$ at which the skyrmions in medium fractionize into half-skyrmions and remains nearly constants in the half-skyrmion phase. This density dependence  in the half-skyrmion
phase indicates that, although $\langle \bar{q}q\rangle \neq 0$ on average in this phase, chiral symmetry is not
restored since hadrons are still massive and there exist pions. In addition, the nearly constant nucleon mass means that it could has a non-vanishing component up to the chiral transition, which might shed light on the origin of the nucleon mass.

}

\FullConference{XV International Conference on Hadron Spectroscopy-Hadron 2013\\
		4-8 November 2013\\
		Nara, Japan }

\begin{document}

\section{Introduction}

The Skyrme model~\cite{Skyrme62} provides a possible natural framework to unifiedly simulate the baryon, baryonic matter and also hadron properties in medium~\cite{PV09,LPMRV03}. From nuclear physics, we learned that the vector mesons, especially the lowest lying ones are essential for understanding the nuclear forces. Based on these considerations, we explored the nuclear matter and medium modified hadron properties by using a chiral effective theory including the pseudoscalar and the lowest lying vector mesons through the hidden local symmetry (HLS) approach~\cite{HY03a}. This talk briefly summarizes what we got in Refs.~\cite{MOYHLPR12,MYOH12,MYOHLPR13,MHLOPR13,Ma:2013vga}.

The following distinctive features of our theoretical framework make our approach differs from the existing literature~\cite{MOYHLPR12,MYOH12}: The chiral effective theory we are using has a self-consistent chiral counting mechanism so that we can control the number of terms to a specific chiral order. In our computation, we used the full Lagrangian written up to the next to leading order of the chiral counting including the homogeneous Wess-Zumino (hWZ) terms which are responsible for the omega meson effect. Thanks to the gauge/gravity duality, with only two input empirical values of the
pion decay constant $f_\pi = 92.4$~MeV and the vector meson mass $m_\rho = 775.5$~MeV, the low  energy constants (LECs) of the effective Lagrangian can be self-consistently determined by a set of master formulas derived from a class of holographic QCD (hQCD) models (here, the Sakai-Sugimoto model~\cite{SS04a,SS05}). Consequently we can make our calculation without any ambiguity and explore the different roles of $\rho$ and $\omega$ mesons in the baryonic matter and hadron properties in medium~\cite{MYOHLPR13,MHLOPR13,Ma:2013vga}.

%In the following, we study the baryonic matter properties and hadron properties in medium using the hidden local symmetry Lagrangian upto %the next to leading order.

\section{Dense baryonic matter from HLS}

\label{sec:matter}

In the sense of large $N_c$ limit, the baryonic matter properties can be simulated by putting the skyrmions from this model onto a crystal lattice, here the face-centered cubic (FCC) crystal, and the baryonic matter density $n$ enters into the system through $n \propto 1/L^3$ with the crystal size $2L$. By varying the crystal size $L$ the baryonic matter density dependence of the relevant quantities can be evaluated (for a review, see, e.g.,~\cite{PV09}). 

\begin{figure}[t]
\centerline{\includegraphics[scale=0.35]{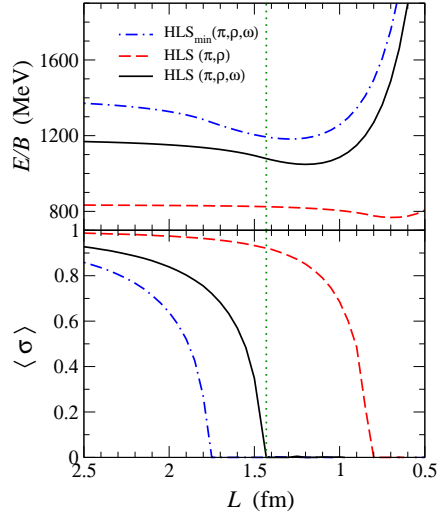}}
%\vspace*{8pt}
\caption{The crystal size $L$ dependence of $E/B$ and $\langle \sigma \rangle$.
 \label{fig:fig1}}
\end{figure}

We show our results of crystal size $L$ dependence of the per-skyrmion energy $E/B$ and
$\langle \sigma\rangle$ with comparisons to other model calculations in Fig.~\ref{fig:fig1}. Here, $\langle \sigma \rangle$ is the space average of the chiral field $U$ over the space that a single skyrmion occupies and it vanishes in the half-skyrmion phase. In the figure, solid, dashed, and dash-dotted lines represent the results of HLS$(\pi,\rho,\omega)$, HLS$(\pi,\rho)$ and HLS$_{\rm min}(\pi,\rho,\omega)$ (for the explicit expression of these models, see Ref.~\cite{MYOH12}), respectively and the vertical dotted line indicates the position of the normal nuclear matter density $n_0$.

From Figure~\ref{fig:fig1} we draw the following lessons of the baryonic matter:
\begin{itemize}

\item

Compared to the results of HLS$_{\rm min}(\pi,\rho,\omega)$, the noticeable observation is that, in HLS$(\pi,\rho,\omega)$, $n_{1/2} \cong n_0$ which indicates that the HLS$(\pi,\rho,\omega)$ describes physics closer to nature. 

\item

The higher value of $n_{1/2}$ from HLS$(\pi,\rho,\omega)$ can be understood from the fact that, compared to HLS$_{\rm min}(\pi,\rho,\omega)$, the single skyrmion size from the former is smaller than that from the later due to the additional interactions in HLS$(\pi, \rho,\omega)$~\cite{MOYHLPR12,MYOH12} which weakens the repulsive interactions from the $\omega$. This understanding is supported by the results from HLS$(\pi,\rho)$, where $n_{1/2}^{} \cong 6n_0$.

\item

The weak dependence of $E/B$ on density in HLS$(\pi,\rho)$ indicates that the inclusion of the $\rho$ meson reduces the soliton mass
and almost saturates the Bogomol'nyi bound.

\end{itemize}

In conclusion, Figure~\ref{fig:fig1} tells us that the vector mesons affect the skyrmion matter properties notably.
The $\rho$ meson favors a high value for the half-skyrmion phase transition
density $n_{1/2}^{}$, while the $\omega$ prefers a smaller value of $n_{1/2}^{}$.
When the both effects are included $n_{1/2}^{}$ becomes similar to the normal
nuclear density $n_0^{}$.

\section{Hadron properties with the FCC crystal background}

\label{sec:hadron}

As we emphasized in above, one of the advantages of the Skyrme model is that we can explore the modified hadron properties in the baryonic matter with this single model. This could be achieved by incorporating the fluctuations of the meson fields with respect to their classical solutions discussed in the previous section, these fields describe the corresponding meson dynamics in dense baryonic matter.

In Ref.~\cite{MYOHLPR13} we studied the medium modified pion decay constant $f_\pi^{\ast}$ by taking into account only the first term of the HLS Lagrangian since at densities away from that of chiral restoration that we are concerning the other terms represent effectively higher order in the chiral counting. In such a case $f_\pi^\ast$ is expressed as
\begin{eqnarray}
\frac{f_\pi^\ast}{f_\pi}
=  \sqrt{1 - \frac{2}{3} \left( 1- \left\langle \sigma^2_{(0)} \right\rangle \right) } .
\label{eq:fpistar}
\end{eqnarray}
Since the space average $\left\langle \sigma^2_{(0)} \right\rangle $ depends on the baryon density through the crystal size $L$, the medium modified $f_\pi^{\ast}$ becomes density dependent.

%%%%%%%%%%%%%%%%%%%%%%%%%%%%%%%%%%%%%%%%%
\begin{figure}[t]
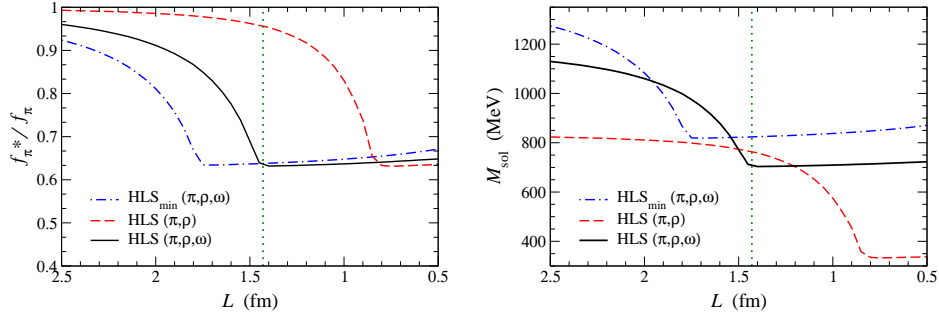

\centering
\includegraphics[scale=0.29]{fig2.eps} \quad
\includegraphics[scale=0.29]{fig3.eps}
\caption[]
{Crystal size dependence of $f_\pi^{\ast}/f_\pi$ and $M_{\rm sol}$ with
the FCC crystal background.
}
\label{fig:fig2}
\end{figure}
%%%%%%%%%%%%%%%%%%%%%%%%%%%%%%%%%%%%%%%%%%

We illustrate $f_\pi^{\ast}/f_\pi$ as a function of the crystal size $L$ in Fig.~\ref{fig:fig2}. From this figure one can easily see that in the single skyrmion and half-skyrmion phases $f_\pi^{\ast}/f_\pi$ behaviors quite differently. In the single skyrmion phase $f_\pi^{\ast}/f_\pi$ decrease monotonically as a function of density while in the half-skyrmion phase it becomes more or less a non-zero constant. Since in the half-skyrmion phase, although the space average $\langle\sigma \rangle \sim \langle\bar{q}q\rangle$ vanishes $f_\pi^*$ does not, the chiral symmetry is still breaking locally, the half-skyrmion phase was interpreted
as a sort of pseudo-gap phase in Ref.~\cite{LPRV03b}.

In addition to $f_\pi^{\ast}$, we also explored the medium modified baryon properties in Ref.~\cite{MYOHLPR13} by substituting $f_\pi^{\ast}$ to the master formulas to determine the medium modified low energy constants. In such a sense, we can explore the medium modified baryon properties and our result of the single skyrmion mass is illustrated in Fig.~\ref{fig:fig2}. This figure shows that the soliton mass $M_{\rm sol}^{\ast}$ has a nearly identical density dependence as $f_\pi^{\ast}$. This feature can be understood by the fact that in the Skyrme model the soliton mass scales roughly as $M_{\rm sol}^{\ast} \sim  f_\pi^{\ast}/e$ with $e$ being the Skyrme parameter, which follows from the scaling of the nucleon mass in the large $N_c$ limit as in matter-free space.

The density dependence of the nucleon mass plotted in Fig.~\ref{fig:fig2} tells us that, in the skyrmion phase with $n < n_{1/2}^{}$, chiral symmetry is broken with the order parameter $\langle \bar{q}q\rangle \neq 0$ and hadrons are massive apart from the Nambu-Goldstone bosons, i.e., the pions, and, in the half-skyrmion phase, we have $\langle \bar{q}q\rangle = 0$ on the average, but chiral symmetry is not restored yet since hadrons are still massive and there
exist pions. The nucleon mass staying as a constant in the half-skyrmion phase might shed light on the
origin of the nucleon mass. What this means is that the nucleon mass could have a component that remains
non-vanishing up to the chiral transition, which is reminiscent of the parity doublet
picture of baryons\cite{DK89}. $M_{\rm sol}^{\ast}$ plotted in Fig.~\ref{fig:fig2} illustrates the crystal size dependence of the part of the nucleon mass originning from the chiral symmetry breaking.

\section{Summary and discussions}
\label{sec:sum}

In this talk, we reported our progress on the structure of baryonic matter based on Skyrme's conjecture. Special emphasize was paid on the effects of $\rho$ and $\omega$ mesons. Our results clearly show that the skyrmion matter properties are
affected by the attractive interaction of the $\rho$ meson and the repulsive interaction
induced by the $\omega$ meson. Explicitly, the incorporation of the tower of isovector vector mesons brings the soliton mass close to the BPS bound while the presence of the isoscalar vector meson $\omega$ obstructs this approach to the BPS state~\cite{Sutcliffe11}.

The simulated hadron properties in the baryonic matter shows that, similar to $f_\pi^{\ast}$, the nucleon mass decreases smoothly as density
increases up to $n_{1/2}^{}$ and, in the half-skyrmion phase, the dropping of
the nucleon mass stops and the mass remains constant going toward to the
chiral restoration. In the skyrmion phase with density $n< n_{1/2}^{}$, chiral symmetry is broken
with the order parameter $\langle \bar{q}q\rangle \neq 0$ and hadrons are
massive except the pions. However, in the half-skrymion phase, even with
$\langle \bar{q}q\rangle = 0$ on the average, chiral symmetry is not restored since
hadrons are still massive and there exist pions.
That the nucleon mass remains non-vanishing as one approaches the chiral transition
point is supported by other approaches. The density dependence of the nucleon mass might shed light on the origin of the nucleon mass.

Finally, we want to say that, some of our results, such as the nucleon mass, the density
$n_{\rm min}^{}$ at which per-baryon mass is at its minimum, the nucleon binding
energy and so on, deviate from nature. These deviations indicate that in our approach some other effects, for example the
dilaton and irreducible three-body interactions, are missing.

%%%%%%%%%%%%%%%%%%%%%%%%%%%%%%%%%%%%%%%%%%

\section*{Acknowledgments}

The work was supported by Grant-in-Aid for Scientific Research on Innovative Areas (No. 2104) ``Quest on New Hadrons with Variety of Flavors'' from MEXT and by the National Science Foundation of China (NSFC) under Grant No.~10905060.

%%%%%%%%%%%%%%%%%%%%%%%%%%%%%%%%%%%%%%%%%%

\end{document}